\documentclass[letterpaper]{article}

\usepackage[T1]{fontenc}

\usepackage{geometry}
\usepackage{setspace}

\usepackage[
  style=chem-acs,
  articletitle=true
]{biblatex}
\usepackage{graphicx}
\usepackage{float}
\newfloat{scheme}{htbp}{los}
\floatname{scheme}{Scheme}
\floatname{chart}{Chart}
\newfloat{graph}{htbp}{loh}
\usepackage{algorithm}
\usepackage{algpseudocode}
\usepackage{chemformula} 
\usepackage[version = 4]{mhchem} 

\usepackage{authblk}
\author[1]{Luis H. Delgado-Granados}
\author[1]{David A. Mazziotti*}
\affil[1]{Department of Chemistry and The James Frank Institute, The University of Chicago, Chicago, IL 60637 USA}

\title{Direct Variational Calculation of Two-Electron Reduced Density Matrices via Semidefinite Machine Learning}
\date{*Email: damazz@uchicago.edu}

\begin{document}

\maketitle

\begin{abstract}
We introduce a data-driven framework for approximating the convex set of $N$-representable two-electron reduced density matrices (2-RDMs). Traditional approaches characterize this set through linear matrix inequalities that define its supporting hyperplanes. Here, we instead learn a vertex-based approximation to its boundary from molecular data and use this information to improve the set defined by low-order positivity constraints, without explicitly constructing higher-order conditions. The resulting semidefinite machine learning approach---combining an input convex neural network with semidefinite programming---drives a direct variational calculation of the 2-RDM with enhanced accuracy at computational cost comparable to two-positivity calculations. Applications to the potential energy curves of ${\rm C}_2^{2-}$, ${\rm N}_2$, and ${\rm O}_2^{2+}$ demonstrate these systematic improvements as well as close agreement with complete active space configuration interaction results. Overall, semidefinite machine learning interweaves data-driven boundary information with semidefinite positivity constraints to yield more accurate energies and 2-RDMs without explicit higher-order positivity conditions.
\end{abstract}


Many-electron systems present a challenging problem due to their computational complexity and unfavorable scaling with system size~\cite{schuch2009computational}. The scaling can be mitigated by replacing the many-electron wave function with the two-electron reduced density matrix (2-RDM)~\cite{M2007, Coleman2000-in}. However, the energy minimization with respect to 2-RDMs---constrained only by Hermiticity, antisymmetry, positivity, and trace---yields energies that are below the ground-state energy~\cite{C1963, garrod1964reduction, mazziotti2012}.  To obtain a valid description of an $N$-electron state, it is necessary to constrain the 2-RDM to be representable by an ensemble $N$-electron density matrix through a set of constraints known as $N$-representability conditions~\cite{C1963, garrod1964reduction, mazziotti2012_physrev}.  By enforcing only a subset of these constraints (e.g., the two-positivity (DQG) conditions), one can generate an accurate lower bound to a molecular ground-state energy through a variational 2-RDM calculation~\cite{mazziotti2001, nakata2001, mazziotti2002, Zhao2004, mazziotti2004, cances2006, shenvi2010, Verstichel.2011, mazziotti2016, li2021,  Knight.2022, mazziotti2023, Gao.2025dgs, schouten2025}.

In this Letter, we propose a physical model based on a combination of machine learning (ML)~\cite{amos2017} and semidefinite programming (SDP)~\cite{boyd1996, mazziotti2011} to directly approximate the $N$-representable set of 2-RDMs~\cite{M2007, Coleman2000-in}.  Traditional constraints tend to characterize $N$-representability through linear matrix inequalities that define the convex set through hyperplanes~\cite{mazziotti2001, nakata2001, mazziotti2002, Zhao2004, mazziotti2004, cances2006, shenvi2010, Verstichel.2011, mazziotti2016, li2021,  Knight.2022, mazziotti2023, Gao.2025dgs, schouten2025}.  Here, we augment this characterization by using data to define a finite set of boundary points that serve as vertices in a polytope approximation~\cite{henk2017basic, Schilling.2013, Liebert.2025axw} of the 2-RDM set, and by using input convex neural networks (ICNNs)~\cite{amos2017, pmlr-v119-makkuva20a, YANG2021107143, wu2024} to learn an approximate boundary of the $N$-representable 2-RDM set through a barrier (penalty) function. The 2-RDM data (e.g., from prior computations) allows us to learn an improved description of the convex 2-RDM set beyond that defined by a set of necessary $N$-representability conditions.  The barrier function is implemented in tandem with an optimization performed by SDP that includes an approximate set of positivity conditions~\cite{mazziotti2001, mazziotti2012_physrev} such as the 2-positivity (DQG) conditions~\cite{C1963, garrod1964reduction}. The characterization of $N$-representability through both vertex-based and hyperplane-based representations—combined through ML and SDP in what we call semidefinite ML—allows us to improve accuracy without significant additional computational cost.  This approach extends earlier work applying machine learning to reduced density matrices~\cite{Schmidt.2021, Sager-Smith.2022, Shao.2023, Jones.2023, delgado-granados2025}, while complementing broader efforts using machine learning to model many-body wave functions~\cite{Carleo.2017, Sajjan.2022} and density functional theory~\cite{Snyder.2012, Moreno.2020}.

To demonstrate the semidefinite ML algorithm with both positivity and ML constraints, we predict the ground-state potential energy curves for a set of triply bonded diatomic molecules: ${\rm C}_2^{2-}$, ${\rm O}_2^{2+}$, and ${\rm N}_2$. The fully $N$-representable 2-RDMs are obtained using complete active space configuration interaction (CASCI)~\cite{Roos.1980, Roos.1987}, while the 2-RDMs with two-positivity conditions are obtained from the variational 2-RDM (v2RDM) method~\cite{mazziotti2001, nakata2001, mazziotti2002, Zhao2004, mazziotti2004, cances2006, shenvi2010, Verstichel.2011, mazziotti2016, li2021,  Knight.2022, mazziotti2023, Gao.2025dgs, schouten2025}.  The semidefinite ML algorithm shows a significant improvement from v2RDM with the 2-positivity conditions, as the potential energy curves obtained are within a few millihartrees of the CASCI results.

\begin{figure}[H]
  \centering
  \includegraphics[width=0.6\linewidth]{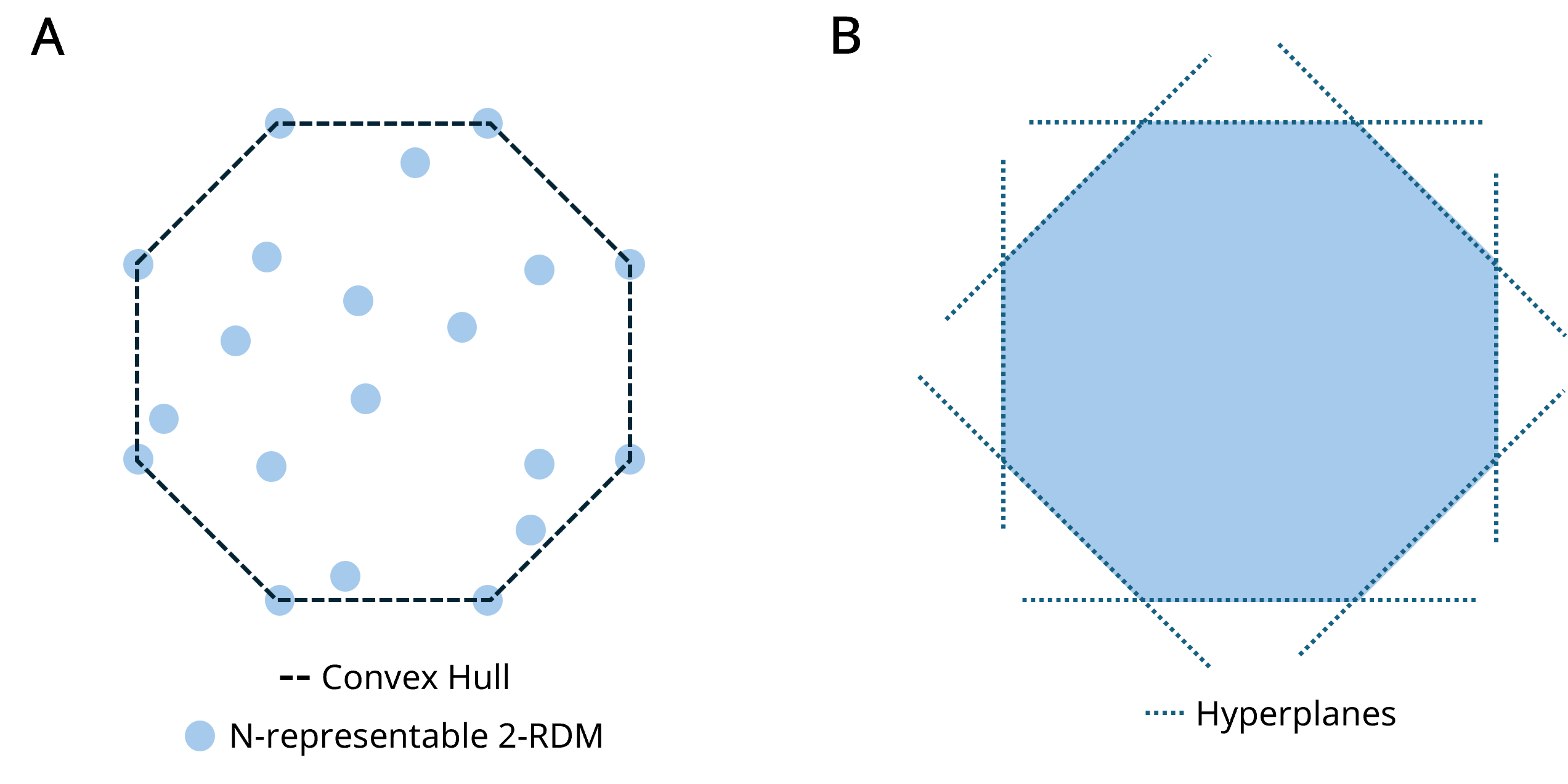} \caption{Representations of a convex set: (A) V-representation and (B) H-representation. While the exact $N$-representable set is not a polytope and therefore does not admit a \emph{finite} vertex representation, the convex hull of a finite set of data-driven boundary (exposed) points provides a V-representation that approximates the true set.}
  \label{fgr:convex-polygon}
\end{figure}

The convex set of $N$-representable 2-RDMs admits two complementary geometric characterizations: (A) a vertex representation (V-representation) as the convex hull of its extreme (boundary) points---that is, the smallest convex set containing those points, formed by all convex combinations of them, and (B) a hyperplane representation (H-representation) as the intersection of hyperplanes that define the set (see Fig.~\ref{fgr:convex-polygon})~\cite{rockafellar1970convex, henk2017basic}. In the V-representation, the set is given by the convex hull of its extreme points. In the H-representation, the same set is characterized by the intersection of hyperplanes that arise from the constraints defining the set.

The implementation of the $N$-representability conditions is generally performed in the H-representation, as the constraints are expressed as a set of linear matrix inequalities that define supporting hyperplanes of the convex set of 2-RDMs~\cite{mazziotti2001, nakata2001, mazziotti2002, Zhao2004, mazziotti2004, cances2006, shenvi2010, Verstichel.2011, mazziotti2016, li2021,  Knight.2022, mazziotti2023, Gao.2025dgs, schouten2025}.  In this work, we use the ICNN~\cite{amos2017} to learn an approximation of the V-representation of the convex set, as the ICNN is constructed to be convex in its inputs (the 2-RDMs).  Each fully $N$-representable 2-RDM obtained from an energetically non-degenerate molecular ground-state calculation corresponds to an exposed boundary point of the convex set~\cite{Mazziotti.1998zdq}, and these data-driven boundary points serve as vertices of a polytope approximation to the true set. The ICNN learns these representative boundary points, which provide information about the contour of the convex hull. In a classification process, the 2-RDMs inside the learned set are assigned a value of zero while the 2-RDMs outside the set are mapped to a nonnegative value that increases with its distance from the convex boundary.

We incorporate the trained ICNN as a barrier function \(\Phi({}^{2}\!D)\) in the energy minimization
\begin{equation}
\begin{aligned}
&\displaystyle{\min_{{}^{2}\!D} \; E\left[{}^{2}\!D\right]+\lambda \Phi({}^{2}\!D)}\\
&\text{s.t.}\quad {}^{2}\!D\in {}^{N}_2\!\widetilde{P},\label{eq:minimization}
\end{aligned}
\end{equation}
where \({}^{N}_2\!\widetilde{P}\) is the convex set of approximately \(N\)-representable 2-RDMs defined by positivity conditions such as DQG~\cite{mazziotti2001, mazziotti2012_physrev, C1963, garrod1964reduction}. The barrier function energetically penalizes 2-RDMs outside the learned \(N\)-representable set, guiding the variational optimization toward the physical region.  Because the energy \(E[{}^{2}\!D]=\mathrm{Tr}({}^{2}\!K\,{}^{2}\!D)\) is linear in \({}^{2}\!D\) and \(\Phi({}^{2}\!D)\) is convex by construction of the ICNN, Eq.~(\ref{eq:minimization}) is a convex optimization problem over a feasible set defined by semidefinite constraints.  Here, ${}^{2}\!K$ is the two-electron reduced Hamiltonian---the matrix representation of the electronic Hamiltonian in the 2-RDM energy functional~\cite{mazziotti2023}.  We minimize this objective using the Frank--Wolfe (conditional gradient) algorithm~\cite{frankwolfe1956}, which exploits the SDP structure of the feasible set.

To implement the minimization, we define the objective
\[
f({}^{2}\!D)=\mathrm{Tr}({}^{2}\!K\,{}^{2}\!D)+\lambda \Phi({}^{2}\!D)
\]
with gradient
\begin{equation}
\nabla f({}^{2}\!D) = {}^{2}\!K + \lambda \nabla \Phi({}^{2}\!D).
\label{eq:grad}
\end{equation}
At iteration \(i\), Frank--Wolfe linearizes the objective about the current iterate \({}^{2}\!D_i\) and computes the search point \({}^{2}\!S_i\) that solves the linear minimization problem
\begin{equation}
\begin{aligned}
{}^{2}\!S_i \in \arg\min_{{}^{2}\!D'\in {}^{N}_2\!\widetilde{P}}
\mathrm{Tr}\!\left(\nabla f({}^{2}\!D_i)\,{}^{2}\!D'\right),
\end{aligned}
\label{eq:minimization_grad}
\end{equation}
which is an SDP because the feasible set is defined by linear matrix inequalities. The next iterate is obtained by a convex combination of the current iterate and the search point,
\begin{equation}
{}^{2}\!D_{i+1}=(1-\omega_i)\,{}^{2}\!D_i+\omega_i\,{}^{2}\!S_i,
\label{eq:convex-comb}
\end{equation}
where the step size \(\omega_i\in[0,1]\) is determined by a line search minimizing
\begin{equation}
\mathrm{Tr}\!\left({}^{2}\!K\,{}^{2}\!D_{i+1}(\omega)\right)
+\lambda \Phi\!\left({}^{2}\!D_{i+1}(\omega)\right).
\label{eq:line-search}
\end{equation}
Iterations continue until convergence, defined by
\[
\|{}^{2}\!D_{i+1}-{}^{2}\!D_i\|_F < \delta.
\]
Geometrically, each iteration identifies the 2-RDM that minimizes an effective Hamiltonian defined by the current gradient, corresponding to an exposed boundary point of the approximate \(N\)-representable set, and updates the solution through a convex combination, progressively approaching the optimal physical 2-RDM. Table~\ref{table:algo} summarizes the resulting semidefinite machine learning algorithm.

\begin{table}[H]
\centering
\caption{Semidefinite ML Algorithm}
\begin{tabular}{l}
\hline
\textbf{Semidefinite ML} \\
\hline
Train ICNN and define barrier function $\Phi({}^{2}\!D)$ \\
Set iteration counter $i \leftarrow 0$ \\
Initialize ${}^{2}\!D_0 \in {}^{N}_2\!\widetilde{P}$ and choose $\lambda$ \\
While $\|{}^{2}\!D_{i+1}-{}^{2}\!D_i\|_F > \delta$: \\
\hspace*{2em} \textbf{Step 1:} Solve linear minimization SDP \\
\hspace*{3em} ${}^{2}\!S_i \in \displaystyle{\arg\min_{{}^{2}\!D' \in {}^{N}_2\!\widetilde{P}} \;
\mathrm{Tr}\{\nabla f({}^{2}\!D_i)\,{}^{2}\!D'\}}$ \\
\hspace*{2em} \textbf{Step 2:} Line search \\
\hspace*{3em} $\omega_i = \displaystyle{\arg\min_{\omega \in [0,1]}
\mathrm{Tr}\!\left({}^{2}\!K\,{}^{2}\!D_{i+1}(\omega)\right)
+\lambda \Phi({}^{2}\!D_{i+1}(\omega))}$ \\
\hspace*{3em} where ${}^{2}\!D_{i+1}(\omega)=(1-\omega){}^{2}\!D_i+\omega{}^{2}\!S_i$ \\
\hspace*{2em} \textbf{Step 3:} Update iterate \\
\hspace*{3em} ${}^{2}\!D_{i+1}=(1-\omega_i){}^{2}\!D_i+\omega_i{}^{2}\!S_i$ \\
\hspace*{2em} \textbf{Step 4:} $i \leftarrow i+1$ \\
\hline
\end{tabular}
\label{table:algo}
\end{table}


Here, we train the ICNN to classify 2-RDMs based on being $N$-representable or not. The training of the ML model is performed with the upper triangular part of the 2-RDMs using a cross-validation scheme (CV). Using only the upper triangular part of the 2-RDM avoids redundant degrees of freedom and allows us to parameterize each 2-RDM by its independent entries; during the SDP optimization, gradients returned by the ICNN are mapped back to the full matrix and explicitly symmetrized to preserve Hermiticity. The output of the ML is a binary representation of being $N$-representable (-1) or not (1). In the case of the barrier function, we employ a piecewise function such that the gradient vanishes once we are inside the $N$-representable set:
\begin{equation}
    \Phi({}^{2}\!D)=\begin{cases}
    0 & \text{if}\quad z < 0 \\
    z^2   & \text{if }\quad z \geq 0
\end{cases}
\quad,
\end{equation}
where $\text{ICNN}({}^{2}\!D)=z$.  Because the ICNN output is convex in its input, $\Phi({}^{2}\!D)=\max(0,z)^2$ is a convex penalty function of ${}^{2}\!D$.  See the method section for additional information on the ICNN, the training process, as well as the feature set.

\begin{figure}[H]
  \centering
  \includegraphics[width=0.6\linewidth]{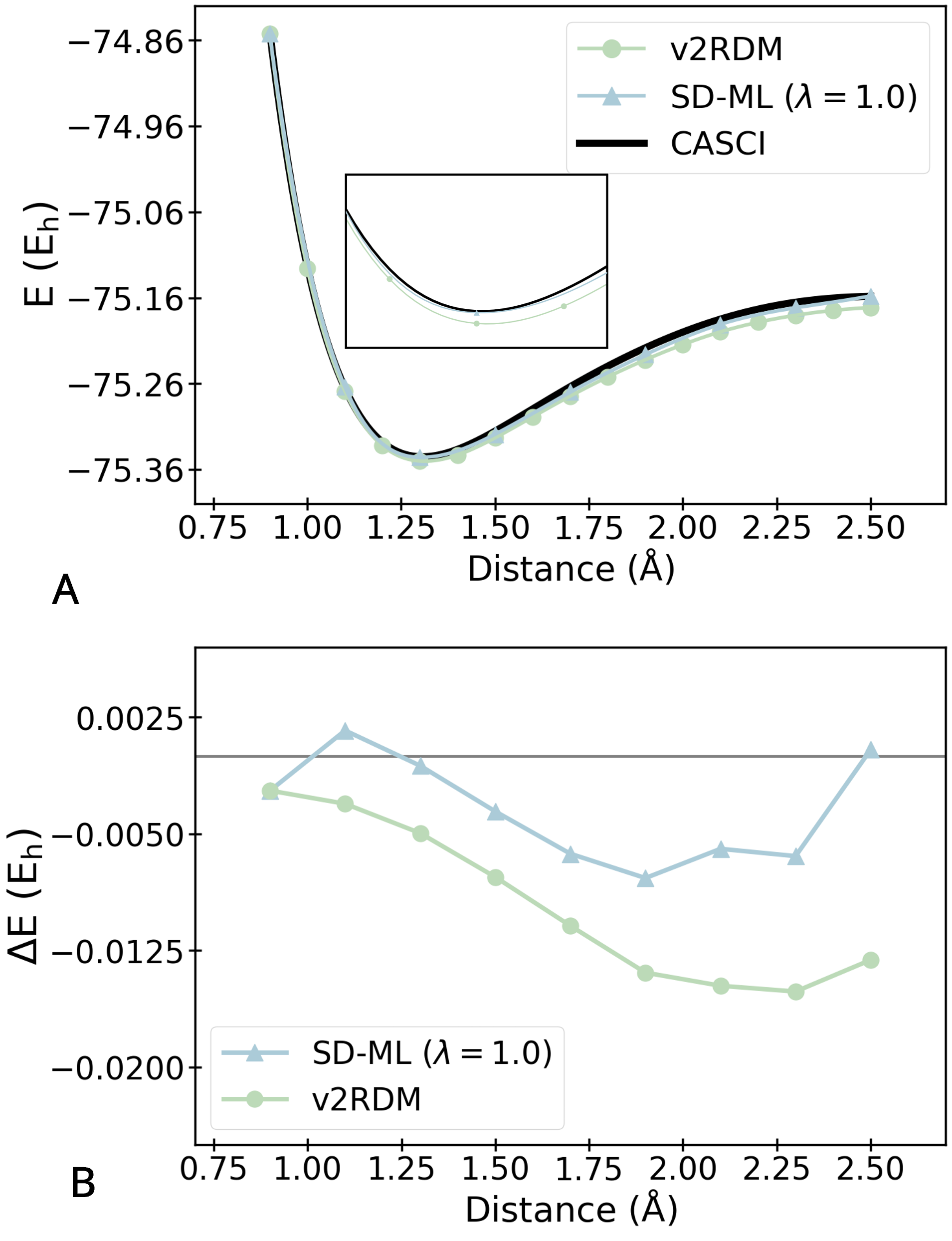} \caption{Predicting ${\rm C}_2^{2-}$ dissociation curve. (a) Variational 2-RDM energies (v2RDM, green circles), complete active space configuration interaction energies (CASCI, black line), and energies predicted by the semidefinite ML algorithm (Table \ref{table:algo}) (SD-ML, blue triangles) are shown as its bond is stretched. (b) Additionally, the errors in v2RDM energies and SD-ML energies compared to CASCI energies are shown. As can be seen, the SD-ML methodology predicts a significantly more accurate potential energy surface for ${\rm C}_2^{2-}$ than v2RDM. Both v2RDM and CASCI energies are computed using the cc-pVDZ basis set and in a $[N_e= 10, N_o= 8]$ active space}
  \label{fgr:C21}
\end{figure}

\begin{figure}[H]
  \centering
  \includegraphics[width=0.6\linewidth]{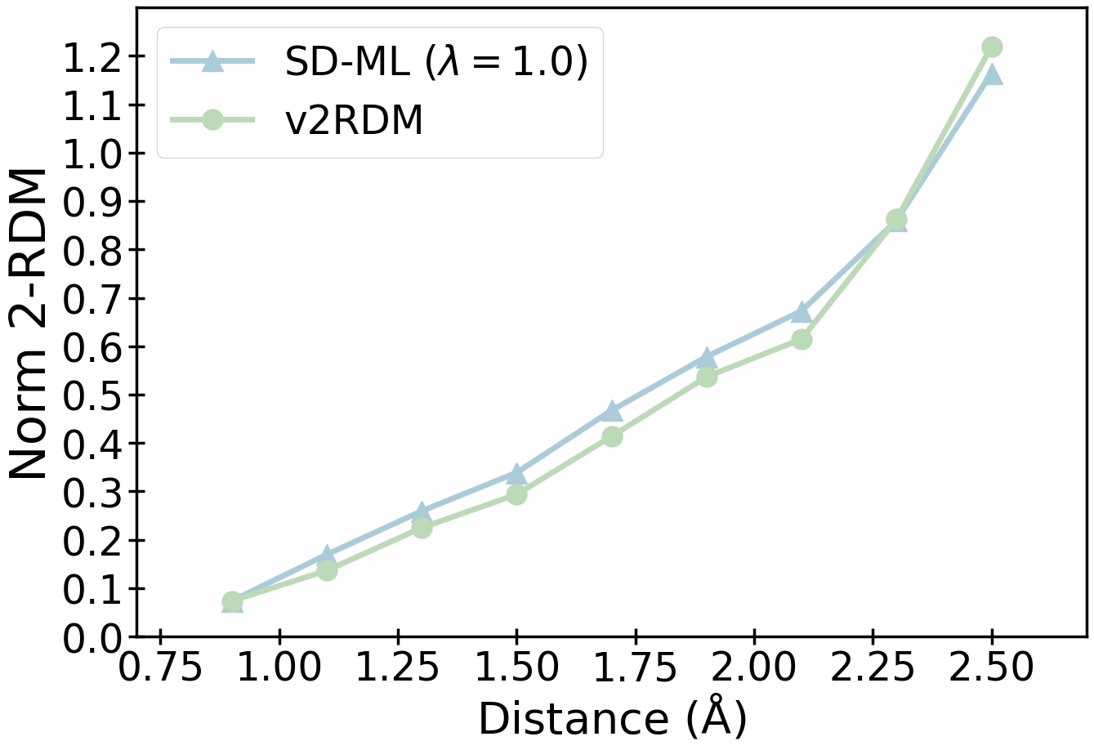} \caption{${\rm C}_2^{2-}$ 2-RDM errors. The Frobenius norms of the difference matrices formed from each v2RDM and SD-ML 2-RDM relative to the CASCI 2-RDM are shown as a function of bond distance. The norms remain similar throughout the bond stretch, with a small improvement at the tail of the potential energy curve.}
  \label{fgr:C21_norm}
\end{figure}

To predict the corrected 2-RDMs for ${\rm C}_2^{2-}$, we compute the 2-RDMs with the 2-positivity conditions at bond distances ranging from $0.9$ to $2.5$~\AA\ using the correlation‑consistent polarized valence double-$\zeta$ (cc-pVDZ) basis set~\cite{dunning1995}. All of the bond distances of the isoelectronic molecules ${\rm O}_2^{2+}$ and ${\rm N}_2$, as well as the even bond distances of ${\rm C}_2^{2-}$ are employed to train the ICNN and define the barrier function $\Phi({}^{2}\!D)$ in the semidefinite ML algorithm described in Table~\ref{table:algo}. As shown in Fig.~\ref{fgr:C21}, the semidefinite ML approach (SD-ML, blue triangles) generates 2-RDMs with energies that are much closer to the CASCI energies (CASCI, black line)  than the v2RDM energies (v2RDM, green circles), demonstrating the ability of this approach to predict 2-RDMs that more accurately represent the state of the system. The only exception to the improvement occurs at 0.9 Å where the ML incorrectly classifies the approximate 2-RDM from v2RDM as $N$-representable.  Because we are using a minimal data approach for illustrative purposes, we expect further improvements with additional training data.
Figure~\ref{fgr:C21_norm} also compares the error in the 2-RDMs from v2RDM and SD-ML relative to the CASCI 2-RDM using the Frobenius norm. The SD-ML and v2RDM norms are similar across the potential curve, with SD-ML showing modest improvement in the stretched, more correlated region of the bond. The Frobenius norm weights each element of the 2-RDM equally, which may mask improvements in key properties such as the energy.

Finally, to explore the generalization of the semidefinite ML, we perform a CV scheme using the three molecule set (${\rm N}_2$, ${\rm C}_2^{2-}$, and ${\rm O}_2^{2+}$), and a feature set analogous to that of ${\rm C}_2^{2-}$. Just as with the previous implementation, we include all the information from the isoelectronic molecules used for training and add the even bond distances of the testing molecule. Results for ${\rm N}_2$ and ${\rm O}_2^{2+}$ show similar improvement in their potential energy curves as ${\rm C}_2^{2-}$. The maximum absolute error in the energy for v2RDM of $20.86$ and $15.25$~mhartree is improved to $7.84$ and $3.22$~mhartree for ${\rm N}_2$ and ${\rm O}_2^{2+}$, respectively. In terms of the Frobenius norm of the 2-RDM error with respect to CASCI, ${\rm N}_2$ and ${\rm O}_2^{2+}$ show similar behavior as ${\rm C}_2^{2-}$, with ${\rm N}_2$ showing a consistent improvement after 1.7~\AA. Further details of the potential energy curves and Frobenius norms for ${\rm N}_2$ and ${\rm O}_2^{2+}$ are provided in the Supporting Information.

In this Letter, we introduce a data-driven framework for approximating the convex set of $N$-representable 2-RDMs. Rather than tightening the feasible set solely by adding linear matrix inequalities, we learn an approximation to its convex boundary and incorporate it as a convex barrier directly within the semidefinite variational optimization. This enables a variational calculation over a substantially improved approximation to the physical set of 2-RDMs while retaining the computational cost of the 2-positivity conditions, without explicit construction of higher-order $N$-representability constraints.  The machine learned barrier function provides a systematic approach to complementing the $N$-representability conditions, expressible as linear matrix inequalities, with data from prior computations about molecular ground states (exposed points) of the 2-RDM set.  While in the present work we separate the hyperplane and vertex information into the SDP and ML parts of the algorithm, respectively, future work will also examine the further synthesis of data with the physics-based constraints.  Although the SD-ML algorithm is presented in a nascent form, it holds promise for providing a general route to data-enhanced, systematically improvable convex formulations of many-electron structure.

\section{Methods}

For all molecules studied, the energies and 2-RDMs are computed using the variational 2-RDM (v2RDM) method~\cite{mazziotti2001, nakata2001, mazziotti2002, Zhao2004, mazziotti2004, cances2006, shenvi2010, Verstichel.2011, mazziotti2016, li2021,  Knight.2022, mazziotti2023, Gao.2025dgs, schouten2025} and the complete active space configuration interaction (CASCI) method~\cite{Roos.1980, Roos.1987} with the cc-pVDZ basis set~\cite{dunning1995}. CASCI calculations for ${\rm N}_2$, ${\rm C}_2^{2-}$, and ${\rm O}_2^{2+}$ are performed in a $[N_e=10,N_o=8]$ active space. These calculations, as well as the SDP implementation, are performed in Maple~\cite{maple_2025} with the Quantum Chemistry Toolbox~\cite{qct_2025,Montgomery2020}.

\section{Machine Learning Algorithm}

\textbf{Model Inputs.} For the feature set, we use the upper triangular 2-RDM.

\noindent \textbf{Model Outputs.} The outputs of the model correspond to a binary representation of being $N$-representable (-1) or not being $N$-representable (1). Because this classification creates a boundary at zero, we employ a squared rectified linear unit (ReLU) function as our barrier function $\Phi$:
\begin{equation}
    \Phi({}^{2}\!D)=\text{max}(0,z)^2,
\end{equation}
where $\text{ICNN}({}^{2}\!D)=z$.

\noindent \textbf{Training \& Testing Data.} The cross-validation scheme is performed on data from ${\rm N}_2$, ${\rm C}_2^{2-}$, and ${\rm O}_2^{2+}$ computed using the cc-pVDZ basis set~\cite{dunning1995}. To obtain the 2-RDMs and potential energy curves, we stretch the bond length of each training molecule across different intervals. The intervals used for ${\rm N}_2$, ${\rm C}_2^{2-}$, and ${\rm O}_2^{2+}$ are $0.9-2.5$ Å,  $0.9-2.5$ Å, and $0.9-2.1$ Å, respectively. When training the model, we include the 2-RDMs corresponding to all geometry points of the molecules that are not being tested. Additionally, we also include the 2-RDMs of even geometry points of the testing molecules, removing them from the testing set. This results in testing being performed only on the odd geometries of the molecule.

\noindent \textbf{Model Specifics.} To build the ICNN, we follow the implementation described by Amos, \textit{et al.} in Ref.~\cite{amos2017} for a fully input convex neural network (FICNN). We use an Adam optimizer, ReLU activation functions in each layer, and a hinge loss function. This neural network is implemented in Python using the API PyTorch~\cite{paszke2019}. In all cases, we use 5-layer FICNN with four layers with $1072$ nodes and a scalar output layer.

\section*{Data Availability}

The data supporting the findings of this study, including potential energy curves and machine learning model parameters, are available from the corresponding author upon reasonable request. These data will also be deposited in a public repository upon publication.

\section*{Acknowledgments}

D.A.M gratefully acknowledges support from the U.S. National Science Foundation Grant No. CHE-2155082.  L.H.D.G. acknowledges that this material is also based upon work supported by the U.S. Department of Energy, Office of Science, Office of Advanced Scientific Computing Research, Department of Energy Computational Science Graduate Fellowship under Award Number DE-SC0024386. This research used resources of the National Energy Research Scientific Computing Center (NERSC), a Department of Energy Office of Science User Facility using NERSC award DDR-ERCAP0026889.

\printbibliography

\end{document}